\documentclass[conference]{IEEEtran}
\IEEEoverridecommandlockouts

\usepackage{cite}
\usepackage{amsmath,amssymb,amsfonts}
\usepackage{algorithm}
\usepackage{algorithmic}
\usepackage{graphicx}
\usepackage{textcomp}
\usepackage{xcolor}
\usepackage{amsmath}
\usepackage{threeparttable}
\usepackage[nonumberlist]{glossaries}
\usepackage{subcaption}
\usepackage{relsize}

\newacronym{1L}{1LDS}{Single Loop Deep Scheduler}
\newacronym{2L}{2LDS}{Double Loop Deep Scheduler}
\newacronym{3GPP}{3GPP}{3rd Generation Partnership Project}
\newacronym{5G}{5G}{Fifth Generation}
\newacronym{5GNR}{5G NR}{5G New Radio}
\newacronym{5GQI}{5GQI}{5G QoS Indicator}
\newacronym{AB}{AB}{Action Branching}
\newacronym{AC}{AC}{Actor-Critic}
\newacronym{ACK}{ACK}{Acknowledgement}
\newacronym{AI}{AI}{Artificial Intelligence}
\newacronym{ATB}{ATB}{Adaptive Transmission Bandwidth}
\newacronym{BLER}{BLER}{Block Error Rate}
\newacronym{BT}{BT}{Bursty Traffic}
\newacronym{BTS}{BTS}{Base Transceiver Station}
\newacronym{C-RNTI}{C-RNTI}{Cell Radio Network Temporary Identifier}
\newacronym{CE}{CE}{Control Element}
\newacronym{CI}{CI}{Confidence Interval}
\newacronym{CLPC}{CLPC}{Closed Loop Power Control}
\newacronym{CNN}{CNN}{Convolutional Neural Network}
\newacronym{CSI}{CSI}{Channel State Information}
\newacronym{CQI}{CQI}{Channel Quality Indicator}
\newacronym{DMRS}{DMRS}{Demodulation Reference Signal}
\newacronym{DQN}{DQN}{Deep Q-Network}
\newacronym{DDQN}{DDQN}{Double Deep Q-Network}
\newacronym{DCI}{DCI}{Downlink Control Information}
\newacronym{DDPG}{DDPG}{DDPG}
\newacronym{DL}{DL}{Downlink}
\newacronym{DL-SCH}{DL-SCH}{Downlink Shared Channel}
\newacronym{DRL}{DRL}{Deep Reinforcement Learning}
\newacronym{DSACD}{DSACD}{Distributional Soft Actor-Critic Discrete}
\newacronym{DRX}{DRX}{Discontinuous Reception}
\newacronym{ERM}{ERM}{Experience Replay Memory}
\newacronym{FB}{FB}{Full Buffer}
\newacronym{FC}{FC}{Fully Connected}
\newacronym{FD}{FD}{Frequency Domain}
\newacronym{FDS}{FDS}{Frequency Domain Scheduling}
\newacronym{FR1}{FR1}{Frequency Range 1}
\newacronym{FTP}{FTP}{File Transfer Protocol}
\newacronym{GAE}{GAE}{Generalized Advantage Estimator}
\newacronym{GBR}{GBR}{Guaranteed Bit Rate}
\newacronym{gNB}{gNB}{gNodeB}
\newacronym{GoB}{GoB}{Grid of Beams}
\newacronym{HARQ}{HARQ}{Hybrid ARQ}
\newacronym{ILLA}{ILLA}{Inner-Loop Link Adaptation}
\newacronym{IRC}{IRC}{Interference Rejection Combining}
\newacronym{JSD}{JSD}{Jensen-Shannon Divergence}
\newacronym{KNN}{KNN}{K-Nearest Neighbors}
\newacronym{KPI}{KPI}{Key Performance Indicator}
\newacronym{L2C}{L2C}{Learning to Communicate}
\newacronym{L2S}{L2S}{Link-to-System-Level}
\newacronym{LCID}{LCID}{Logical Channel ID}
\newacronym{LMMSE}{LMMSE}{Linear Minimum Mean Square Error}
\newacronym{LLR}{LLR}{Log-Likelihood Ratio}
\newacronym{LLS}{LLS}{Link-Level Simulator}
\newacronym{LS}{LS}{Least Squares}
\newacronym{LTE}{LTE}{Long Term Evolution}
\newacronym{MAC}{MAC}{Medium Access Control}
\newacronym{MARL}{MARL}{Multi-Agent Reinforcement Learning}
\newacronym{MCS}{MCS}{Modulation and Coding Scheme}
\newacronym{MCTS}{MCTS}{Monte Carlo Tree Search}
\newacronym{MIMO}{MIMO}{Multiple Input Multiple Output}
\newacronym{ML}{ML}{Machine Learning}
\newacronym{MMIB}{MMIB}{Mean Mutual Information per Coded Bit}
\newacronym{MNO}{MNO}{Mobile Network Operator}
\newacronym{MRC}{MRC}{Maximal-ratio combining}
\newacronym{MSE}{MSE}{Mean Square Error}
\newacronym{MUMIMO}{MU-MIMO}{Multi-User MIMO}
\newacronym{NACK}{NACK}{Negative Acknowledgement}
\newacronym{OFDM}{OFDM}{Orthogonal Frequency-Division Multiplexing}
\newacronym{OLPC}{OLPC}{Open Loop Power Control}
\newacronym{OLLA}{OLLA}{Outer-Loop Link Adaptation}
\newacronym{PDB}{PDB}{Packet Delay Budget}
\newacronym{PDCCH}{PDCCH}{Physical Downlink Control Channel}
\newacronym{PDCP}{PDCP}{Packet Data Convergence Protocol}
\newacronym{PDSCH}{PDSCH}{Physical Downlink Shared Channel}
\newacronym{PDU}{PDU}{Protocol Data Unit}
\newacronym{PF}{PF}{Proportional Fair}
\newacronym{PG}{PG}{Policy Gradient}
\newacronym{PHY}{PHY}{Physical Layer}
\newacronym{PMI}{PMI}{Precoder Matrix Indicator}
\newacronym{PPO}{PPO}{Proximal Policy Optimization}
\newacronym{PUCCH}{PUCCH}{Physical Uplink Control Channel}
\newacronym{PUSCH}{PUSCH}{Physical Uplink Shared Channel}
\newacronym{QoS}{QoS}{Quality of Service}
\newacronym{RACH}{RACH}{Random Access Channel}
\newacronym{RAN}{RAN}{Radio Access Network}
\newacronym{RAR}{RAR}{Random Access Response}
\newacronym{RE}{RE}{Resource Element}
\newacronym{RB}{RB}{Resource Block}
\newacronym{RBG}{RBG}{Resource Block Group}
\newacronym{RI}{RI}{Rank Indicator}
\newacronym{RL}{RL}{Reinforcement Learning}
\newacronym{RLC}{RLC}{Radio Link Control}
\newacronym{RR}{RR}{Round Robin}
\newacronym{RRC}{RRC}{Radio Resource Control}
\newacronym{RSRP}{RSRP}{Reference Signal Received Power}
\newacronym{RZF}{RZF}{Regularized Zero Forcing}
\newacronym{SAC}{SAC}{Soft Actor-Critic}
\newacronym{SACD}{SACD}{Soft Actor-Critic Discrete}
\newacronym{SAW}{SAW}{Stop-And-Wait}
\newacronym{SD}{SD}{Spatial Domain}
\newacronym{SDS}{SDS}{Spatial Domain Scheduling}
\newacronym{SE}{SE}{Spectral Efficiency}
\newacronym{SEM}{SEM}{Standard Error of the Mean}
\newacronym{SL}{SL}{Single Loop}
\newacronym{SoTA}{SoTA}{State-of-the-Art}
\newacronym{SSSG}{SSSG}{Search Space Set Group}
\newacronym{SDU}{SDU}{Service Data Unit}
\newacronym{SINR}{SINR}{Signal-to-Interference-and-Noise Ratio}
\newacronym{SLS}{SLS}{System-Level Simulator}
\newacronym{SNR}{SNR}{Signal-to-noise Ratio}
\newacronym{SS}{SS}{Synchronization Signal}
\newacronym{SUMIMO}{SU-MIMO}{Single-User MIMO}
\newacronym{TB}{TB}{Transport Block}
\newacronym{TD}{TD}{Time Domain}
\newacronym{TBS}{TBS}{Transport Block Size}
\newacronym{TDD}{TDD}{Time Division Duplexing}
\newacronym{TDS}{TDS}{Time Domain Scheduling}
\newacronym{TPC}{TPC}{Transmit Power Control}
\newacronym{TS}{TS}{Technical Specification}
\newacronym{TTI}{TTI}{Transmission Time Interval}
\newacronym{UCI}{UCI}{Uplink Control Information}
\newacronym{UE}{UE}{User Equipment}
\newacronym{UPT}{UPT}{User Perceived Throughput}
\newacronym{UL}{UL}{Uplink}
\newacronym{UL-SCH}{UL-SCH}{Uplink Shared Channel}
\newacronym{XR}{XR}{Extended Reality}
\newacronym{VR}{VR}{Virtual Reality}
\newacronym{vRAN}{vRAN}{virtualized Radio Access Network}
\newacronym{ZF}{ZF}{Zero Forcing}

\IEEEaftertitletext{\vspace{-1.5\baselineskip}}

\title{Combining Improvements in Uplink AI-RAN}

\author{\IEEEauthorblockN{
Petteri Kela}
\IEEEauthorblockA{\textit{Nokia Technologies} \\
Espoo, Finland}
\and
\IEEEauthorblockN{
Dani Korpi}
\IEEEauthorblockA{\textit{Nokia Bell Labs} \\
Espoo, Finland}
\and
\IEEEauthorblockN{
Mikko Honkala}
\IEEEauthorblockA{\textit{Nokia Bell Labs} \\
Espoo, Finland}
}

\begin{document}

\renewcommand{\baselinestretch}{1}

\maketitle

\begin{abstract}
One of the major transformative factors in 6G will be the integration of \gls{AI} to become a native part of \gls{RAN}. While most physical-layer AI features have so far been evaluated in isolation using link-level simulations, their combined behavior in a realistic multi-cell, multi-UE deployment has remained largely unexplored. In this paper, we present system-level performance results when multiple uplink AI features are enabled together, achieved by integrating accurate link-level and system-level simulators. To infer state-of-the-art deep-learning-aided \gls{MIMO} receivers under the dynamic allocations produced by a realistic uplink scheduler, we propose a mirrored data augmentation method that decouples receiver performance from scheduled allocation size. In addition to these \gls{PHY} receiver features, we combine several recent advances in deep reinforcement learning to train uplink power control and link adaptation that outperform a heuristic baseline and further boost the gains obtainable from the \gls{AI} receiver alone. The system-level results show that the combined \gls{AI} features improve the mean uplink user throughput by roughly 27\% compared to a non-\gls{AI} baseline, confirming that the individual \gls{PHY} and \gls{MAC} \gls{AI} features provide complementary gains when deployed jointly.
\end{abstract}


%%%%%%%%%%%%%%%%%%%%%%%%%%%%%%%%%%%%%%%%%%%%%
\section{Introduction}

\gls{AI} and \gls{ML} are emerging as key enablers to improve the performance of next-generation 6G \gls{RAN}. It was already envisioned in \cite{9040431} that deep learning would eventually replace \gls{PHY} and \gls{MAC} layer algorithms, and \cite{10597082} surveys existing studies on embedding individual ML features into the MAC layer. However, these features have so far been developed and evaluated largely in isolation — typically at link-level for \gls{PHY} features and under simplified assumptions for \gls{MAC} features — leaving their combined behavior in a realistic multi-cell, multi-\gls{UE} deployment largely unexplored. This work closes that gap.

One of the most prominent \gls{PHY} \gls{AI} features is the fully-learned receiver. DeepRx \cite{deeprx1,deepRx_2_5} is a \gls{CNN}-based receiver that replaces the conventional channel-estimation, equalization, and demapping chain with a single deep model. By learning data-aided detection, it achieves high performance even with considerably sparser \gls{DMRS} patterns than conventional receivers require, and it scales to high-order \gls{MIMO} scenarios. Owing to its convolutional structure, DeepRx can in principle operate over varying bandwidth allocations. However, such AI receivers have so far been evaluated almost exclusively at link-level with fixed \gls{RB} allocations, whereas a realistic dynamic uplink \gls{MIMO} scheduler assigns allocations that rarely match the fixed size the model was trained on, and the resulting allocation-edge discontinuities corrupt the detected \glspl{LLR}. To bridge this gap, we propose mirrored data augmentation, which reflects and concatenates the received frequency-domain samples to synthetically fill the trained \gls{RB} allocation — giving the scheduler full freedom without retraining or an allocation-specific model zoo.

Both link adaptation and power control have recently been studied with ML and are considered well-suited to \gls{RL} \cite{11358408, 10597082, 10024770, 9410444}, since their per-\gls{TTI} decisions make training tractable with sample-efficient \gls{DRL}. We target the inner-loop \gls{MCS} selection (\gls{ILLA}), where a learned policy can better handle imperfect \gls{SINR} estimates, while retaining classical \gls{OLLA} to track the \gls{BLER} target.

Power control, in turn, employs both open-loop and closed-loop mechanisms to adjust the uplink transmission power of \glspl{UE}. The open-loop parameters — normalized transmit power density $P_0$ and pathloss compensation factor $\alpha$ — set the base transmission power, which is then dynamically refined through closed-loop \gls{TPC} commands. \gls{TPC} commands are signaled either as absolute values or as accumulated adjustments. The accumulated option is more attractive for \gls{ML} purposes, since it can raise and lower a UE's transmission power dynamically and thereby form an arbitrary power distribution across \glspl{UE}. Some prior works apply \gls{ML} to \gls{CLPC}, such as \cite{9410444}, which demonstrates the gain potential of \gls{ML}-based \gls{CLPC} for a single UE per cell. However, realistic deployments require further study with multiple multiplexed \glspl{UE} and cells.

In this work, we integrate several state-of-the-art deep-learning-based features into our industry-grade \gls{SLS}, which combines accurate link-level and system-level modeling. We first describe our deep learning features for \gls{UL} channel estimation, receiver, power control, and link adaptation. We then show, through extensive system-level simulations with multiple \glspl{UE} and cells and realistic dynamic frequency-domain scheduling, how these features contribute combined gains and, when combined, outperform a heuristic baseline by a large margin. To the best of our knowledge, this is the first work to evaluate these \gls{PHY} and \gls{MAC} \gls{AI} features jointly in a realistic multi-cell, multi-\gls{UE} system-level setting, rather than in isolation at either link or system-level. The remainder of the paper is organized as follows: Section II presents the DeepRx receiver features, Section III describes the \gls{DRL}-based DeepLA and DeepTPC features, Section IV details the simulation methodology and numerical results, and Section V concludes the paper.

%%%%%%%%%%%%%%%%%%%%%%%%%%%%%%%%%%%%%%%%%%%%%
\section{DeepRx Features}

As the \gls{PHY} receiver algorithm, we consider DeepRx \cite{deeprx1,deepRx_2_5}. It is a \gls{CNN}-based \gls{AI} receiver, which demonstrates robust performance under a wide range of operating conditions. In particular, it is capable of using the unknown signal constellation points to further refine the soft bit estimates, which provides significant performance gains under imprecise channel estimates.
\begin{figure}[htb]
  \centering
  \includegraphics[width=\columnwidth, trim={0cm 14.0cm 18.5cm 0.3cm}, clip]{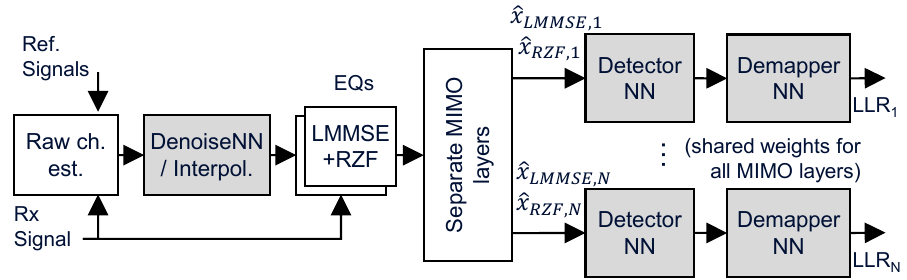}
  \caption{The considered DeepRx AI receiver model architecture~\cite{deepRx_2_5}.}
  \label{fig:deeprx}
  \vspace{-2mm}
\end{figure}

The latest variant in \cite{deepRx_2_5} represents the state-of-the-art of AI receivers under inter-cell-interference and high mobility. It builds on deep \glspl{CNN} in two key parts of the receiver chain: channel estimate denoiser and post-equalization signal correction. The complete receiver architecture is depicted in Fig.~\ref{fig:deeprx}, while the deep learning -based elements are described below in more detail. Note that in this paper we refer to the receiver as DeepRx, for simplicity, although in \cite{deepRx_2_5} it is named EqDeepRx to differentiate from prior AI model variants.

\subsection{DeepRx Denoiser}

In order to improve the accuracy of the channel estimates, the DMRS-based raw channel estimates are processed with \gls{CNN}-based DenoiseNN. It is fed with the channel estimate of each pilot-carrying \gls{RE}, and processes them to provide the denoised estimates at the output. The denoised estimates are then interpolated along the frequency- and time-direction as per the usual procedure. Note that the DenoiseNN operates in parallel for each Tx-Rx antenna pair, meaning that the spatial dimension is not considered. This ensures highly efficient parallelization in high-order MIMO systems. For further details, please refer to \cite{deepRx_2_5}.

\subsection{DeepRx Detector}

The denoised channel estimate is used for \gls{IRC}-type equalization, similar to a legacy \gls{LMMSE} receiver. However, DeepRx incorporates additional \gls{ML}-based post processing in the form of a DetectorNN, which operates on the equalized soft symbols. To extract the most performance gains from the DetectorNN step, an additional \gls{RZF}-equalization is performed in parallel to the \gls{LMMSE}-\gls{IRC} equalizer. DetectorNN processes these inputs and outputs soft symbol estimates for each \gls{RE}, which are then fed to the demapper. In this work, an \gls{ML}-based demapper is used, although similar performance can be obtained also with a conventional non-ML demapper. It should be noted that also DetectorNN operates independently per each MIMO stream, or layer. This ensures that the complexity scales favorably with increasing spatial multiplexing order, albeit at the cost of being unable to obtain additional performance gains from mitigating cross-stream interference. Again, we refer the reader to \cite{deepRx_2_5} for additional details.

\subsection{Mirrored Data Augmentation}

We noticed that our pretrained DeepRx models did not perform as expected straight out of the box with the dynamic \gls{FD} scheduler such as \gls{ATB} \cite{4657148}, which optimizes the power spectral density between frequency domain multiplexed \glspl{UE}. Varying contiguous allocation sizes became a problem for DeepRx models trained only for contiguous and fixed 16 \gls{RB} blocks. Since inference runs on a fixed 16-RB × 14-symbol block, a narrower allocation leaves the rest of the block empty. Those zeros are not merely uninformative: the convolutional receptive field spans the whole block, so the step discontinuity at the allocation edges propagates inwards and corrupts the \glspl{LLR} of the allocated \glspl{RB}.

In order to resolve this, we do not retrain per allocation size nor with varying allocation sizes, but instead make the network input resemble the fully occupied block it was trained on. This is done by filling the block using mirrored data augmentation. The band is partitioned into fixed 16-\gls{RB} blocks, each processed independently, so a block may be only partially covered by the allocation. The covered part is reflected about both of its edges and, when it is narrow, extended by alternating reversed and forward tiles until the block is full; the same reflection is applied to the received signal and to the channel estimate so that the padded region stays a consistent observation–channel pair. The key property is boundary continuity — the sample just outside the allocation equals the nearest allocated one — unlike cyclic replication, which restarts the pattern at every tile seam and injects artificial frequency-domain discontinuities. For simplicity, only \glspl{LLR} of the allocated \glspl{RB} are decoded. However, the reflected images, which are independent network outputs for the same bits, may optionally be combined by \gls{MMIB} weighting.

While DeepRx with zero padding or cyclic repetition did not provide gain over the \gls{LMMSE}-\gls{IRC} baseline with dynamic scheduling, mirrored data augmentation restores near-training performance under dynamic \gls{FD} scheduling, without a retraining or allocation-size-specific model zoo.

%%%%%%%%%%%%%%%%%%%%%%%%%%%%%%%%%%%%%%%%%%%%%
\section{Deep Reinforcement Learning Features}

In addition to the \gls{PHY} receiver AI features, we use deep reinforcement learning to learn \gls{ILLA} and \gls{CLPC}. During the rest of the paper, we will refer to these features as DeepLA and DeepTPC, respectively.

As a \gls{DRL} algorithm, we use the \gls{SACD} variant \cite{christodoulou-sacd}, designed specifically for discrete action spaces, while the original \gls{SAC} \cite{haarnoja2018soft} shines more with continuous action spaces. \gls{SAC} variants are especially known for their sample efficiency, making them a good weapon of choice for learning \gls{RAN} features. \gls{SACD} variant, used in this paper, already showed solid learning performance in other \gls{RAN} features such as \gls{DL} \gls{MUMIMO} scheduling \cite{kela2025from} and dynamic duplexing~\cite{11563648}.

\subsection{Enhancements to \gls{SACD}}
In order to further improve performance, inspired by \cite{hessel2018rainbow}, we combined further advancements in reinforcement learning into \gls{SACD}. These features include recycling of dormant neurons, prioritized replay memory, connection skipping, and multi-step learning. The list of final tuned hyper-parameters and enhancement features used with DeepLA and DeepTPC are given in Table \ref{tab_rl_hyper}.

\subsubsection{Recycling of Dormant Neurons}
To avoid inactive neurons during training \cite{ReDo} proposes the ReDo algorithm that computes an activation score for each neuron based on its average activation magnitude. Neurons whose score falls below a threshold are considered dormant. All weights entering the dormant neuron are reset using the same random initialization distribution used at the start of training. This gives the neuron a new representation to learn. All weights from the dormant neuron to the next layer are set to zero. This ensures that reactivating the neuron does not suddenly disturb the current network output. With DeepTPC this provided a small improvement, but DeepLA did not benefit notably from ReDo.

\subsubsection{Critic Connection Skipping}
\cite{sinha2021drl} proposes a simple approach, in which the original input is concatenated into every hidden layer of the policy and value networks through dense skip connections, named D2RL. As stated in \cite{sinha2021drl}, applying D2RL to Q networks of \gls{SAC} generally provides a greater benefit than having it only in policy. The practical reason for applying D2RL to only Q networks is also that those are needed only during training. Hence, a real time operating system in a base station may use just the policy network of \gls{SACD} and avoid the added computational complexity of D2RL in real time inference. This feature was found to be beneficial for training both DeepLA and DeepTPC.

\subsubsection{Prioritized Replay}
Prioritized replay and multi-step learning were considered the most important enhancements in \cite{hessel2018rainbow}. Instead of sampling randomly from the replay memory, the prioritized experience replay
samples transitions with probability relative to the last encountered absolute TD error. The shape of the probability distribution is shaped by the hyper-parameter prioritization exponent $\omega$.

\subsubsection{Multi-Step Learning}
We implement multi-step bootstrapping for \gls{SACD} as described in \cite{11563648}. Hence, instead of using a single reward corresponding to state $s$ and next state $s+1$ soft state-value function expectation, we use two-step discounted reward with expectation from state $s+2$. For the DeepLA we did not apply multi-step learning due to maximizing instantaneous throughput with zero discount. However, for the DeepTPC this feature was the key factor for successful learning, because increasing power step-by-step too much with accumulated \gls{TPC} commands backfires as increased interference when neighbor cells' start to increase their \glspl{UE}' powers as well.

\subsection{DeepLA}
The goal of DeepLA is simply to replace traditional \gls{ILLA} and predict uplink \gls{MCS} that provides the best throughput under the upcoming channel conditions. Thus, the action space is \gls{MCS} indices. The input vector includes the following eight features from allocated \glspl{RB}: (i) mean \gls{SINR}, (ii) minimum \gls{SINR}, (iii) maximum \gls{SINR}, (iv) standard deviation of \gls{SINR}, (v) number of allocated \glspl{RB}, (vi) mean \gls{SINR} delta compared to past average\gls{SINR}, (vii) number of \gls{MIMO} streams, and (viii) number of consecutive \glspl{NACK}. These inputs are normalized to be within the range of $[-1,1]$.
The output is a softmax probability vector including probability for each \gls{MCS}.
As a reward, we use the throughput per \gls{RB} maximizing function:
\begin{equation}
r = \begin{cases}
\dfrac{\text{TB}_{\text{kbits}}}{ N_{\text{RB}}}, & \text{ACK} \\
-0.1, & \text{NACK}
\end{cases},
\end{equation}
where ${TB}_{\text{kbits}}$ denotes the transport block size in kbits and $N_{\text{RB}}$ is the allocation size in \glspl{RB}. A small negative reward is given from \glspl{NACK} to avoid too many failures. During training \gls{SACD} samples actions based on softmax probabilities, but once the model is trained, we use argmax for the best performance.

\subsection{DeepTPC}
With DeepTPC we learn to select the accumulated \gls{TPC} commands that network signals to \glspl{UE} via \gls{DCI}. The \gls{5G} standard assumes a 2-bit \gls{TPC} command table corresponding to $[-1, 0, 1, 3]$, which defines our action space. Because \gls{TPC} commands are prone to power increase, we omit the 3 dB increase and use only a subset which is $[-1, 0, 1]$ dB accumulations. In the end, \gls{SACD} would learn to adjust power with the full \gls{TPC} command table, but the balanced action space avoids drifting towards maximum powers in the beginning and helps \gls{SACD} training when the model is exploring. In order to avoid too much variance, we also limit the maximum accumulation. We noticed that without limitations, we could get even higher mean user throughput gains, but that would backfire slightly at the cell edge. Hence, optimal geometric mean user throughput gain, which emphasizes proportional fairness, was obtained with an accumulation range of $[-6, 6]$ dBs.

The DeepTPC input feature vector consists of four features including, received SINR, current Tx power with accumulation, number of allocated \glspl{RB} and \gls{RSRP}. The reward function aims at optimizing the received effective \gls{SINR} as follows:
\begin{equation}
r =
\begin{cases}
-10, & \text{SINR}_{\text{rx}} < -6 \text{dB} \\
\operatorname{clip}_{[-1,1]}\!\left(\frac{\text{SINR}_{\text{rx}} - \text{SINR}_{\text{target}}}{10} + 1\right), & \text{otherwise}
\end{cases},
\end{equation}
where
\begin{equation}
\begin{split}
\text{SINR}_{\text{target}} = P_0 &+ 10\log_{10}f_{\text{scs}} + 10\log_{10}N_{\text{rx}} \\
&- (1-\alpha)\text{PL} - P_{I+N}
\end{split}.
\end{equation}
The target \gls{SINR} is calculated using \gls{5G}'s Tx power \gls{OLPC} formula \cite{3GPPTS38213} using normalized transmit power density $P_0$, subcarrier spacing factor $f_{\text{scs}}$, number of receiver antennas $N_{rx}$, pathloss $\text{PL}$ scaled with pathloss compensation factor $\alpha$, and estimated interference plus noise power $P_{I+N}$. The reward aims at providing the maximum reward if received $\text{SINR}_\text{rx} \ge \text{SINR}_\text{target}$. In order to avoid hurting the cell edge \glspl{UE} we use a negative $-10$ reward for small effective $\text{SINR}_\text{rx}<-6 \text{dB}$ that start to become hard to decode.

After training, instead of switching to argmax, we noticed that with DeepTPC it is beneficial to keep some entropy alive and sample \gls{TPC} commands from the policy network's output keeping Tx power levels in constant movement. However, after training, we reduce the entropy by taking the softmax function out of the equation to make the magnitude of highest sampling weights larger compared to lowest ones. Thus, some entropy remains, but it is significantly reduced for ratio-based sampling.
    \begin{table}[t]
        \caption{SACD Hyper-parameters}
        \label{tab_rl_hyper}
        \centering
        \begin{threeparttable}
        \begin{tabular}{lll}
            \hline
            Parameter & DeepLA & DeepTPC\\ \hline
            Learning rate & 0.0003 & 0.0001\\ 
            Weight initialization & uniform [-0.2, 0.2] & uniform [-0.2, 0.2]\\
            Discount factor &  0 & 0.95\\ 
            Target entropy C &  0.5 & 0.9 \\ 
            Mini-batch size &  64 & 64 \\ 
            Hidden layers &  2 & 2 \\ 
            Input size &  8 & 4 \\ 
            Output size &  28 & 3 \\ 
            Neurons per hidden layer \tnote{*} & 64 & 64 \\
            Replay prioritization $\omega$ & 0.5 & 0.5 \\
            Optimizer algorithm & Adam & Adam\\
            Activation function & ReLU & ReLU\\
            Target soft update rate & 0.01 & 0.01\\
            ReDo & disabled & enabled \\
            Critic D2RL & enabled & enabled \\
            Multi-step learning & disabled & 2 steps \\
            Training action sampling & softmax & softmax \\
                                            & weighted random & weighted random \\
            Inference action sampling & argmax & weighted random \\
            \hline
        \end{tabular}
        \begin{tablenotes} \footnotesize \item[*]D2RL adds input amount of neurons to 
        critics' hidden layers. \end{tablenotes} \end{threeparttable}
    \end{table}

%%%%%%%%%%%%%%%%%%%%%%%%%%%%%%%%%%%%%%%%%%%%%
\section{Performance Evaluation}

\subsection{Combining Link and System-Level Simulation}
As already envisioned in \cite{10482909}, to study the system-level performance of ML-enabled air interface solutions, the traditional look-up table based \gls{L2S} interface has to be replaced with the proper \gls{LLS} approach, which can accurately simulate different channel estimation and receiver methods with intra-slot channel aging. In particular, the Sionna open source library \cite{Sionna} is used for this \gls{LLS} part. \gls{LLS} provides \glspl{LLR} for industry grade \gls{SLS}, which uses the well known \gls{MMIB} method \cite{MMIB} to calculate effective \gls{SINR} for each transmission \cite{10482909}. In this way, we can simulate the link-level at the \gls{OFDM} symbol level, while \gls{SLS} runs with granularity of the slots. A summary of simulation parameters is provided in Table \ref{tab_sim_params}.
    \begin{table}[t]
        \caption{Simulation configuration.}
        \centering
        \begin{tabular}{ll}
            \hline
            Parameter & Value \\ \hline
            Deployment scenario & 3GPP Urban Macro, 500m ISD \cite{3GPPTR38913}, \\ 
                                & 3D channel model in \cite{3GPPTR38901} \\
            Number of BSs       & $21$ (7 sites, 3 sectors/site) \\
            Number of UEs       & $210$  \\ 
            UE velocity         & 30 km/h \\
            BS antenna config & (M, N, P, Mg, Ng, Mp, Np) = \\
                              & (8, 2, 2, 1, 1, 1, 2) \\ 
                              & 32 antenna elements with 4 TXRUs \\
            UE antenna config & 4 port realistic directive antennas as in \cite{3GPPTR38901}: \\
                              & 10 \% UE without blockage, \\
                              & 52.2 \% UE with one hand grip, \\
                              & 26.1 \% of UE with two hand grips, \\
                              & 11.7 \% of UE with head blockage \\
            BS Receiver & LMMSE-IRC,  \\
                        & DeepRx Denoiser with LMMSE-IRC, or \\
                        & DeepRx Denoiser with DeepRx Detector \\
            BS channel estimation & \gls{LS} with Gaussian smoothing \\
                                  & (used with all receiver options) \\
            DMRS configuration & Type 1, 1 additional position \cite{3GPPTS38211} \\
            MIMO & single-user MIMO \\
                 & Type 1 single-panel UL codebook \cite{3GPPTS38214} \\
            Transmission rank & adaptive (1-2) \\
            Carrier frequency & FR1, $4$ GHz \\ 
            System bandwidth & $20$ MHz (50 RBs) \\
            Duplexing & TDD (DDDDU) \\
            PHY settings & $30$ kHz subcarrier spacing, \\
                         & 12 subcarriers per RB \\
                         & 14 OFDM symbols per slot \\ 
            BS noise figure & $5$ dB \\
            Traffic model & Full Buffer, or \\
                          & FTP Model 3 (2 Mbps with 0.25 MB packets) \\
            Packet scheduler & Adaptive transmission bandwidth \cite{4657148} \\
            Uplink power control & UE max Tx power: 26 dBm \\
                                 & ${P}_{0}$: -86 dBm, $\alpha$: 0.8 \\
            Link adaptation & ILLA, or DeepLA \\
                            & OLLA with BLER target of 10\% \\
                            & MCS Index Table 2 for PUSCH \cite{3GPPTS38214}\\
            HARQ & Asynchronous incremental redundancy \\
                 & with max 4 retransmissions \\
            Number of drops & 30 in evaluations, 1 in RL model training \\
            \hline
        \end{tabular}
        \label{tab_sim_params}
    \end{table}

\subsection{Deep Learning Model Training}

DeepRx models are pretrained in Sionna environment \cite{Sionna} as described in \cite{deepRx_2_5}. The \gls{DRL} models are pretrained using the simulation environment defined in Table \ref{tab_sim_params} with a single random drop, before evaluating the models within different random simulation realizations. We use a centralized learning entity that receives training samples from each cell to its replay memory. This helps us to train generalized models that can be used in all base stations. We first train the DeepLA model with DeepRx features enabled. During DeepLA training, the full buffer traffic model is enabled and \gls{OLLA} is disabled in order to learn throughput maximizing \glspl{MCS} selections. Once DeepLA is trained, we train DeepTPC with all other \glspl{ML} features enabled. In order to obtain results with realistic \gls{BLER} target, we use \gls{OLLA} with 10\% BLER target. DeepLA can outperform the legacy link adaptation by a greater margin if \gls{OLLA} is disabled, but then \gls{BLER} can become higher and occasionally cause additional delays due to undesired radio link failures.

\begin{figure}[!t]
    \setlength\abovecaptionskip{0.0\baselineskip}
    \centering
    \includegraphics[width=\columnwidth, trim={0.5cm 0cm 0.1cm 0cm}, clip]{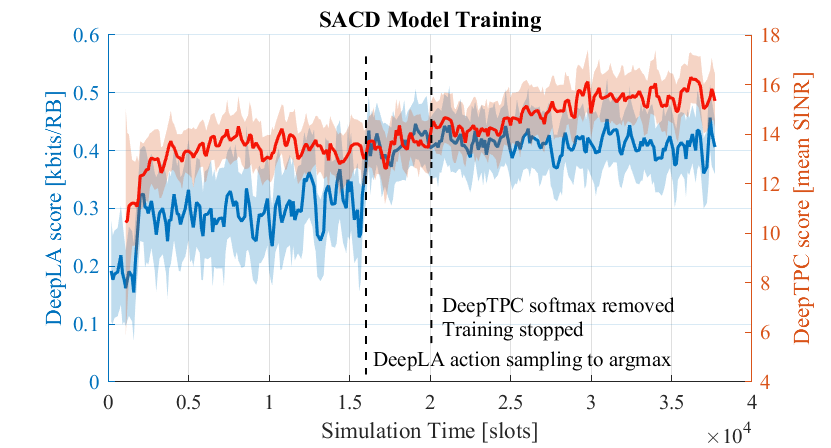}
    \caption{Mean performance scores of DeepLA and DeepTPC training with $\pm2\sigma$ range across base stations. DeepLA's performance is visualized with kbits/RB scoring while DeepTPC's performance score is effective SINR driven.}
    \label{fig_drl_training}
    \end{figure}

Fig. \ref{fig_drl_training} illustrates the effective \gls{SINR} evolution during DeepTPC training and the evolution of kbits/RB during DeepLA training runs. As discussed earlier, trained DeepLA benefits from switching to argmax-based action selection, while DeepTPC benefits from entropy reduction to keep power levels varying slightly. As shown in Fig. \ref{fig_drl_training}, we train both models $20\,000$ slots collecting training samples from all transmissions of 210 \glspl{UE}. DeepLA is switched to argmax action selection a moment before the training period is stopped to allow all other network algorithms to converge before results collection starts. The remainder of the training simulation runs are used to evaluate that the trained models work as expected.

We also tried training DeepTPC and DeepLA simultaneously. Policies trained like that also outperformed their baselines, but slightly better outcomes were observed by training separately due to less noise in rewards.

\subsection{Numerical Results}

    \begin{figure*}[!ht]
    \setlength\abovecaptionskip{0.5\baselineskip}
    \centering
    \begin{subfigure}[t]{0.49\textwidth}
      \centering
      \includegraphics[width=\columnwidth, trim={0.7cm 0cm 0.7cm 0cm}, clip]{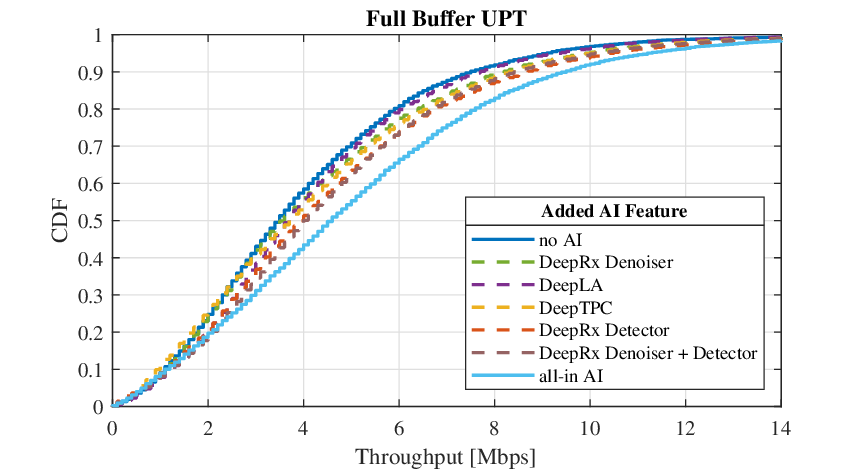}
      \caption{CDF of the full-buffer \gls{UPT}.}
      \label{fig_add}
    \end{subfigure}
    \hfill
    \begin{subfigure}[t]{0.49\textwidth}
      \centering
      \includegraphics[width=\columnwidth, trim={0.7cm 0cm 0.7cm 0cm}, clip]{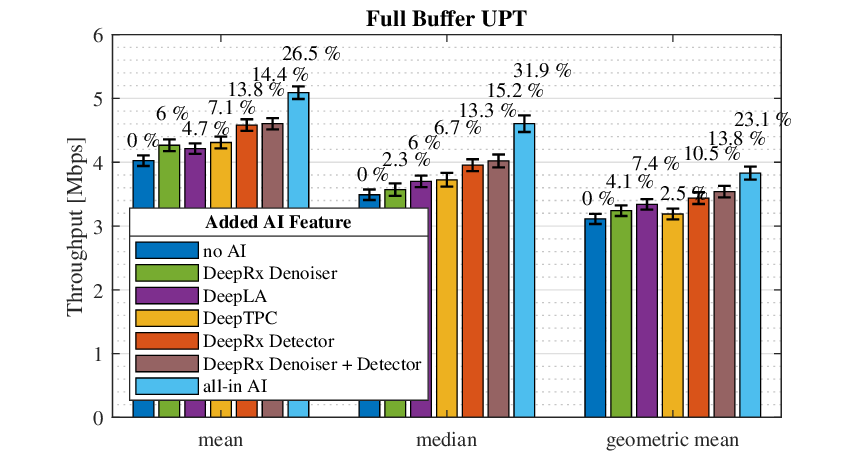}
      \caption{Mean, median, and geometric mean \gls{UPT} gains, with 95\% \gls{CI} error bars over 30 random drops.}
      \label{fig_add_bars}
    \end{subfigure}
    \caption{Full-buffer uplink \gls{UPT} when individual \gls{AI} features are enabled on top of the no-\gls{AI} baseline (features are not cumulatively stacked), except all-in \gls{AI}, which enables all features jointly. The DeepRx Detector has the single largest individual impact. Note that DeepLA and DeepTPC are trained to be used with DeepRx, while here used with plain LMMSE-IRC.}
    \label{fig_individual}
    \vspace{-3mm}
    \end{figure*}
    
    \begin{figure*}[!th]
    \setlength\abovecaptionskip{0.5\baselineskip}
    \centering
    \begin{subfigure}[t]{0.49\textwidth}
      \centering
      \includegraphics[width=\columnwidth, trim={0.7cm 0cm 0.7cm 0cm}, clip]{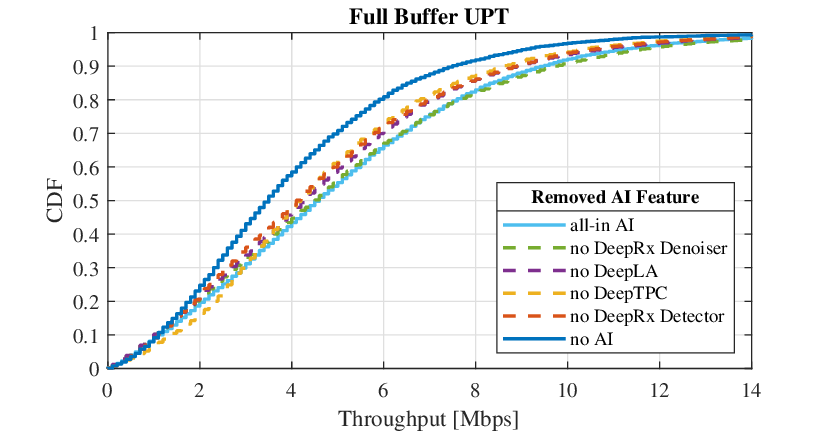}
      \caption{CDF of the full-buffer \gls{UPT}, with the curves ordered by feature significance.}
      \label{fig_remove}
    \end{subfigure}
    \hfill
    \begin{subfigure}[t]{0.49\textwidth}
      \centering
      \includegraphics[width=\columnwidth, trim={0.7cm 0cm 0.7cm 0cm}, clip]{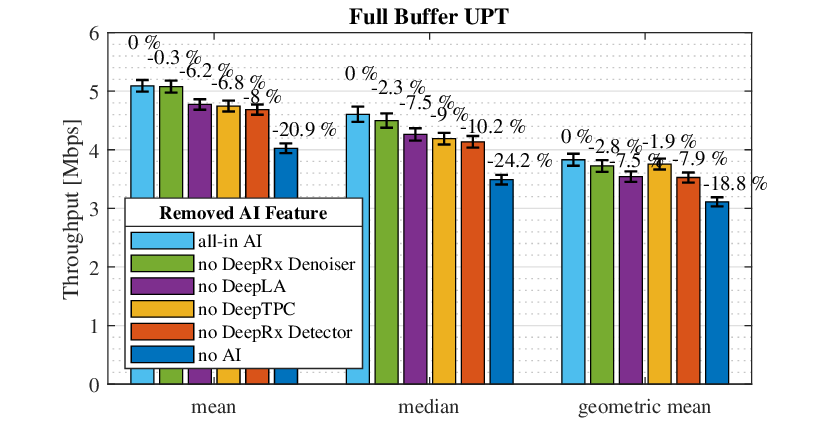}
      \caption{Mean, median, and geometric mean \gls{UPT} loss per removed feature. Error bars: 95\% \gls{CI} over 30 drops.}
      \label{fig_remove_bars}
    \end{subfigure}
    \caption{Full-buffer uplink \gls{UPT} under feature ablation, where one \gls{AI} feature at a time is removed from the all-in \gls{AI} configuration. Comparing against the all-in \gls{AI} and no-\gls{AI} bounds isolates the marginal contribution of each feature. The DeepRx Detector is the most significant feature and partially compensates for the removal of the DeepRx Denoiser; viewed in this light, the latter proves to be the least significant one.}
    \label{fig_ablation}
    \vspace{-3mm}
    \end{figure*}

For numerical results, we deploy our pretrained models to simulations with multiple different random seeds and traffic models without further training. Although DeepTPC and DeepLA are trained with DeepRx features enabled, we first show the performance increase provided by each individual AI feature over no AI option in Figs. \ref{fig_add} and \ref{fig_add_bars}. It can be seen that the all-in AI option outperforms other options, as expected, providing a mean throughput gain of 27\%. If different AI features are individually compared, then it can be seen that the DeepRx Detector has the biggest positive impact on \glspl{UPT}. Although DeepLA and DeepTPC are trained with both DeepRx features enabled, gains also exist with the baseline \gls{LMMSE}-\gls{IRC} receiver. This indicates good generalization by the respective models.
    
Because DeepTPC and DeepLA are trained with DeepRx features enabled, it makes sense to look at the loss of ablation that every single AI feature provides in Figs. \ref{fig_remove} and \ref{fig_remove_bars}. Because, DeepRx Denoiser and Detector can partially compensate the absence of each other the removal of one does not provide as much performance drop as would be expected based on Fig. \ref{fig_add_bars}. Despite that, the DeepRx Detector remains the most significant \gls{AI} feature. These observations are in line with the original link-level studies in \cite{deepRx_2_5}. Moreover, now we can better see the significant gain contribution of DeepLA and DeepTPC, when these models are used with the features they were trained with.

\begin{figure*}[!t]
    \setlength\abovecaptionskip{0.5\baselineskip}
    \centering
    \begin{subfigure}[t]{0.49\textwidth}
      \centering
      \includegraphics[width=\columnwidth, trim={0.7cm 0cm 0.7cm 0cm}, clip]{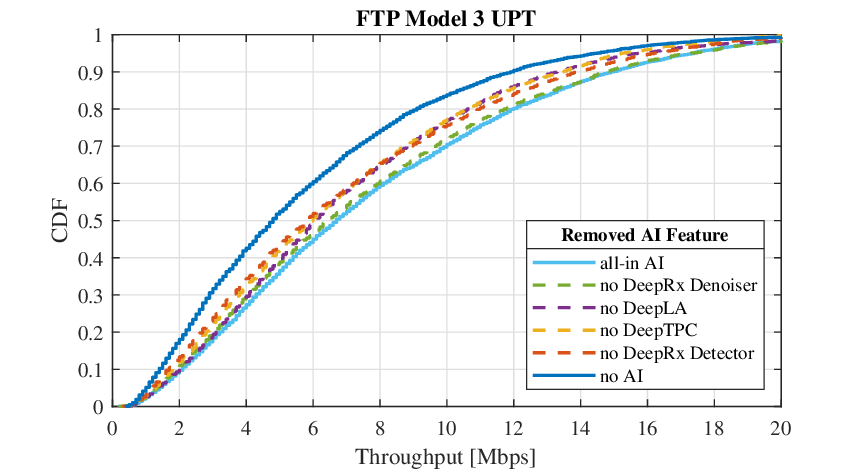}
      \caption{CDF of the \gls{UPT} with FTP Model 3 traffic.}
      \label{fig_ftp3_remove}
    \end{subfigure}
    \hfill
    \begin{subfigure}[t]{0.49\textwidth}
      \centering
      \includegraphics[width=\columnwidth, trim={0.7cm 0cm 0.7cm 0cm}, clip]{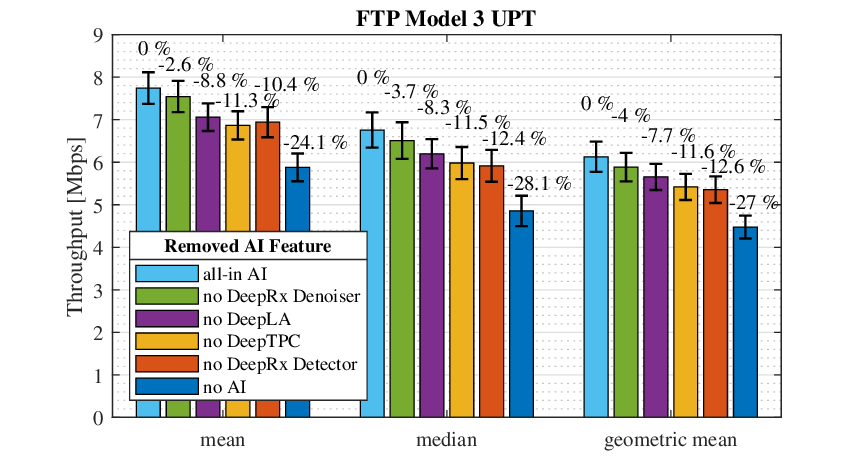}
      \caption{Mean, median, and geometric mean \gls{UPT} with FTP Model 3 traffic. Error bars: 95\% \gls{CI} over 30 drops.}
      \label{fig_ftp3_remove_bars}
    \end{subfigure}
    \caption{User perceived throughput with FTP Model 3 traffic. The importance of the \gls{AI} features remains, or even slightly increases, with bursty traffic at $\sim50$\% \gls{RB} load. Removing all \gls{AI} features results in a 24\% drop in mean throughput, which corresponds to the 32\% benefit provided by \gls{AI}.}
    \label{fig_ftp3}
    \vspace{-3mm}
    \end{figure*}

In addition to full buffer traffic, we have experimented with the same pre-trained AI models with FTP Model 3 traffic model with parameterization that creates roughly 50\% \gls{RB} usage. Figs. \ref{fig_ftp3_remove} and \ref{fig_ftp3_remove_bars}, which show AI feature ablation comparison, indicate that the same models also work with bursty traffic and improve \glspl{UPT} across the \gls{UE} distribution in all random drop simulation realizations.

As a summary of numerical results with realistic \gls{SLS}, we observe that the DeepRx performance gains are largely in line with the \gls{LLS} results reported in \cite{deepRx_2_5}. More importantly, our \gls{SLS} results show that combining DeepRx with DeepLA and DeepTPC yields cumulative throughput gains which far exceed the gain achievable by any individual AI model. With improved receiver performance, power control has more freedom to lower transmission power to reduce interference or increase it to deliver more signal power — whichever yields the best overall trade-off. DeepTPC learns this per \gls{UE}; the scheduler can assign wider RB allocations because \glspl{UE} are not as easily power limited, and link adaptation can assign higher \gls{MCS} while still reaching the \gls{BLER} target. By learning \gls{ILLA} with DeepLA, \gls{OLLA} operates closer to the optimal point, improving performance. Although the uplink power control parameters ($\alpha$ and $P_0$) were exhaustively searched for their optimal values, DeepTPC was able to further improve performance by learning a better power distribution between all \glspl{UE} and finding a better balance between interference and received power.

\subsection{Computational Complexity}

As described in \cite{deepRx_2_5}, DeepRx complexity and \gls{BLER} performance can be traded off through architectural hyperparameters such as network width and depth. As complexity optimization was not our focus, we conservatively opted for a configuration somewhat larger than \cite{deepRx_2_5}, with DenoiseNN modified to use three residual blocks with 128 channels instead of the reference configuration using four residual blocks with 64 filters. We also modified DemapperNN to use 64 filters instead of 32 as in \cite{deepRx_2_5} and increased its' depth from 3 to 4 resnet blocks.  All other architectural components follow \cite{deepRx_2_5} unchanged. The resulting implementation contains 219k trainable parameters and requires approximately 0.68~GFLOPs per inference per 16~PRBs (192 subcarriers) per MIMO layer. Of the total complexity, DenoiseNN accounts for 57k parameters and 0.09~GFLOPs, while DetectorNN and DemapperNN together account for 162k parameters and 0.59~GFLOPs. The overall complexity of the model is therefore roughly twice that of the primary EqDeepRx configuration reported in \cite{deepRx_2_5}.

The inference latencies of the DeepLA and DeepTPC \gls{SACD} policy networks were measured on an Intel\textsuperscript{\textregistered} Xeon\textsuperscript{\textregistered} Gold 6330 CPU (2.0 GHz). A proprietary C++ implementation using FP32 precision achieved single-thread inference latencies below 2 $\mu$s per decision. Since link adaptation and transmit-power-control decisions are made at slot timescales (hundreds of microseconds), the measured inference overhead is negligible relative to the scheduler execution budget. These results indicate that DeepLA and DeepTPC can be deployed in real-time base-station software without dedicated \gls{ML} inference accelerators.

%%%%%%%%%%%%%%%%%%%%%%%%%%%%%%%%%%%%%%%%%%%%%
\section{Conclusions}

In this paper, we demonstrated how multiple state-of-the-art deep-learning features for the uplink air interface can be combined and evaluated together in a realistic, industry-grade System-Level Simulator (SLS) that integrates accurate link-level modeling. By coupling a Link-Level Simulator (LLS) with an SLS through the \gls{MMIB} effective \gls{SINR} interface, we were able to capture effects — most notably inter-cell pilot contamination and dynamic frequency-domain scheduling — that link-level studies alone cannot reproduce. To enable inference of AI receivers under the dynamic uplink allocations produced by a realistic scheduler, we introduced mirrored data augmentation, and we combined several recent reinforcement learning advancements (dormant-neuron recycling, critic connection skipping, prioritized replay, and multi-step learning) into the \gls{SACD} algorithm used to train DeepLA and DeepTPC.

Our results show that these AI features deliver mutually reinforcing gains and, when enabled together, outperform a strong heuristic baseline by a large margin. The all-in AI configuration achieved uplink throughput gains of roughly 27\% in the mean, 32\% in the median, and 23\% in the geometric mean over the no-AI baseline. Among individual features, the DeepRx Detector had the single largest impact. This is a key observations, as it confirms that the previously demonstrated link-level performance gains of DeepRx also translate to system-level gains. Another important observation is that the DRL-based MAC features provided substantial additional gains on top of the DeepRx AI receiver: DeepLA moved ILLA closer to its optimal operating point, while DeepTPC learned a better inter-UE power distribution and improved the balance between interference and received power — despite the open-loop parameters $P_0$ and $\alpha$ having already been exhaustively optimized.

Future work includes deeper integration of the AI algorithms, system-level evaluations under a broader range of conditions, and the inclusion of additional AI features.

\bibliographystyle{IEEEtran}
\bibliography{bib}

\end{document}